\def\mb{M_{\rm b}}
\def\rhoc{\rho_{\rm c}}
\def\rhob{\rho_{\rm B}}

\def\req{R_{\rm eq}}
\def\msol{M_\odot}
\def\lcrit{\lambda_{\rm crit}}
\def\lmax{\lambda_{\rm max}}

\def\Ener{\cal E}
\def\Press{p}
\def\nb{n_{\rm b}}
\def\nA{n_{\rm A}}
\def\nB{n_{\rm B}}

\def\gamA{\Gamma_{\rm A}}
\def\gamB{\Gamma_{\rm B}}
\def\gamm{\Gamma_{\rm m}}
\def\m0{m_{0}}
\def\mm{m_{_{0{\rm m}}}}
\def\mA{m_{_{0{\rm A}}}}
\def\mB{m_{_{0{\rm B}}}}
\def\KA{K_{\rm A}}
\def\KB{K_{\rm B}}
\def\Km{K_{\rm m}}

\documentclass[]{aa}
\usepackage{graphicx}
\usepackage{natbib}
\usepackage[usenames]{color}
\usepackage{multirow}

\begin{document}
\topmargin -1.6in
\title{Phase transitions in rotating neutron stars cores: back bending,
stability, corequakes, and pulsar timing}
\author{J. L. Zdunik\inst{1}
\and
 M. Bejger\inst{1}
 \and
 P. Haensel\inst{1} \and
 E. Gourgoulhon\inst{2}
 }
\institute{N. Copernicus Astronomical Center, Polish
           Academy of Sciences, Bartycka 18, PL-00-716 Warszawa, Poland
{\em jlz@camk.edu.pl}}
\institute{N. Copernicus Astronomical
Center, Polish
           Academy of Sciences, Bartycka 18, PL-00-716 Warszawa, Poland
\and LUTH, UMR 8102 du CNRS, Observatoire de Paris, F-92195 Meudon Cedex, France\\
{\tt jlz@camk.edu.pl, bejger@camk.edu.pl,  haensel@camk.edu.pl,
  Eric.Gourgoulhon@obspm.fr}}
\offprints{J.L. Zdunik}
\date{Received: 28.09.2005, Accepted: 06.01.2006}
\abstract{}{We analyze potentially observable phenomena during spin
evolution of isolated pulsars, such as back bending and  corequakes
  resulting from instabilities, which could
 result from  phase transitions in neutron star cores.}
{We study these aspects of spin evolution of isolated compact stars
by means of analytical  models of equations of state, for
both constant-pressure phase transitions and the transitions through
the mixed-phase region. We use high-precision 2-D multi-domain
spectral code LORENE for the calculation of the   evolutionary
sequences of rotating neutron stars. This allows us to  search the
parameter space for possible instability regions, and possible
changes in the stability character of rotating stars with phase
transitions in their cores.} {We determine the conditions on the
density jump in constant-pressure phase transitions which leads
to the back bending phenomena or to the existence of
the unstable segments in the evolutionary sequences of spinning down
isolated normal neutron stars.
We formulate the conjectures concerning
the existence of two disjoint families of non-rotating and rotating
stationary configurations of neutron stars.
To clarify the effect of rotation on the stability of neutron star we
present the particular case of EOSs leading to marginal instability of static
and rotating configurations: marginal instability
point in non-rotating configurations continues to exist in all
evolutionary spin-down tracks.
We discuss the fate of rotating stars entering
the region of instability calculating the change in radius, energy
release, and spin-up associated with the corequake in rotating
neutron star, triggered by the instability.
The energy release is found to be very  weakly dependent on the angular
momentum of collapsing star.}{}


\titlerunning{Rotating NSs and phase transitions}

\maketitle

%
\section{Introduction}
\label{sec.introd}
%
The properties of the high-density matter in compact stars
are now studied by means both  of nuclear physics and astrophysics.
Many different equations of state (EOS) were
proposed in the literature, based on different
theoretical models of dense matter. In particular,
it has been postulated that at high enough density,
matter contains hyperons, and that phase transitions
to exotic phases of dense matter, such as meson (pion, kaon)
condensates, or quark matter can
occur in the dense compact star cores (for a recent review of
possible states of matter in compact star cores see e.g.
\citealt{glendbook97}). Unfortunately, the lack of precise knowledge
of strong interactions between hadrons in dense matter, as well
as deficiencies and approximations plaguing the many-body calculations
of the EOS prevent us from knowing
the actual structure of neutron star cores. Terrestrial experiments
cannot supply information on the properties of matter
at density  $\ga  10^{15}~{\rm g~cm^{-3}}$ expected
at neutron star center.  We can only hope that observations
of neutron stars will provide us with constraints which
will enable us to select a correct dense matter model, or at
least to limit the number of acceptable dense matter theories.

Particularly interesting method of searching for the phase
transition  in neutron star cores via  pulsar timing was proposed
by \citet{GPW97}. As a pulsar spins down, its central density
increases, and  for a certain density a new phase of matter can
appear. In the case considered   by \citeauthor{GPW97},  the new
phase consisted of quark matter. The authors suggested, that the
softening of the EOS, induced by the formation of the new dense
phase, leads to a temporary spin-up era, the phenomenon called
{\it back-bending}. Originally, the name  comes from nuclear physics,
where the  phenomenon of ``back-bending'' was observed in the
systematics of the moment of inertia of excited states of rapidly
rotating nuclei, see e.g., \citet{rs80}. The calculations of
\citeauthor{GPW97}
 were performed within the slow-rotation approximation
 (\citealt{hartle67,ht68}), supplemented with additional
 relations  resulting from accounting for the  rotational
 stretching and frame-dragging  effects \citep{wg91,wg92}.

Several other authors \citep{hh98,CGPB00} carried out
their calculations of the back-bending phenomenon using the
slow rotation approximation of Hartle. However, as shown by
\citet{sbgh94}, the Hartle method,
when  compared with results of {\it exact} 2-D numerical codes,
  breaks down for angular velocity close to the Keplerian one.

First calculations concerning back-bending based on {\it exact}
2-D code were performed by \citet{cheng02}. These authors
used the version of KEH code \citep{keh89a,keh89b}, improved
by \citet{sf95} (see also references therein). In their work,
\citeauthor{cheng02} focused on the role of the crust
for the very  existence of the back-bending. Indeed, as they
show, even a slight change in the physical state of the crust
(for example, a change in the crust-core transition pressure)
may significantly affect the results. This shows that  high
precision is mandatory for reliable calculation of the
back-bending phenomenon.

Another important article containing results based on 2-D
computations with the
\citet{sf95} code, was published by \citet{nick02}. They
showed that the results obtained by \citet{GPW97} are
plagued  by large numerical uncertainties. For example,
the very same EOS as that used by \citeauthor{GPW97}
(EOS from Table 9.2 of \citealt{glendbook97}) did not yield
the back-bending phenomenon at all in \citet{nick02}!
It became evident that the back-bending problem is
much subtler than previously considered, and that it
requires careful handling as well as high-precision
2-D computations.
Also, \citeauthor{nick02} pointed out some errors in
previous papers on back-bending. For example, the formula for
braking index must be corrected by taking into account
the rotational flattening of the star (Sect. 6 in \citealt{nick02}).

Most recently, \citet{zhgb04} showed that the back-bending
phenomenon  can also occur for the EOSs different from the
mixed-phase one. They also pointed out importance of the
stability with respect to the axi-symmetric  perturbations.
The appearance of hyperons in the dense matter
\citep{bg97} softens some EOSs so much,
that pulsars  losing angular momentum  actually spin up
during a  period of time. Paradoxically, during this
spin-up phase pulsars could lose a significant
amount of their angular momentum.

It should be stressed that all previous works (except of
\citealt{zhgb04}) considered back-bending as a feature of the
the dependence of the moment of inertia, $I$, on the
rotation frequency, $f\equiv 1/{\rm rotation~period}$:
  $I=I(f)$. As shown recently by \citet{zhgb04}, this can easily
lead to incorrect determination of the stability of rotating
stars. \citet{zhgb04} pointed out that many cases claimed before
to correspond to the back-bending, actually cannot be realized in
nature because of the instability with respect to the
axi-symmetric perturbations.

One of the aims of the present work is to determine reliably and
precisely  the stability regions on the back bending segments of
the spin evolution tracks. In order to avoid any precision
problem, and to investigate large and possibly complete parameter
space, we will work with analytical  EOSs of dense matter
exhibiting a softening at supra-nuclear density. Two examples of
softening by a phase transition will be considered. We will
study EOSs with constant-pressure phase transitions, characterized
by a density jump obtained using the Maxwell construction. Second,
we will use EOSs with phase transition extending over a finite
pressure range in which two pure phases coexist forming a
mixed-phase state. Such EOSs are obtained for the first-order
phase transition between two pure phase by relaxing the condition
of local electric charge neutrality and replacing it by less
stringent condition of the global neutrality (\citealt{glend92}).
Mixed-phase state can be realized provided the surface tension at
the interface between the two pure phase is not too large.

The plan of the article is as follows. In  Sect.\ \ref{sec.eos} we
present various types of analytical  EOSs used in the calculation
of the spin evolution tracks. Numerical methods used in exact 2-D
calculations are briefly presented in Sect.\ \ref{sec.numerical}.
Our numerical results are described in Sect.\
\ref{sec.back-bending}. We first describe the general criteria for the
back-bending and the stability for spinning-down stars. Then, the results
for the EOSs with a mixed-phase segment are studied in Sect. \
\ref{sec.mixed}, and those with constant pressure phase transition
with density jump are reviewed in Sect.\ \ref{sec.maxwell}. In
Sect.\ \ref{sec.conjecture} we describe a link between the
existence of unstable segments in the families of static and
rotating configurations of neutron stars. Change in neutron star
parameters, accompanying transitions between two rotating
configurations, triggered by instabilities of isolated rotating
neutron stars,
 are  studied in Sect.\ \ref{sect:corequake}. Modifications in the
 pulsar timing and pulsar age evaluations, due to phase transitions
 in spinning-down isolated neutron stars are
 studied in Sect.\ \ref{sect:timing.age}.
  A summary of our results and  their discussion is presented  in final
 Sect.\ \ref{sect:conclusions}. Formulae referring to
 analytical models of EOSs with phase transitions are
 collected in the Appendix.
%
\section{Analytical EOS}
\label{sec.eos}
%

\begin{figure}[h]
\centering \resizebox{3.5in}{!}{\includegraphics{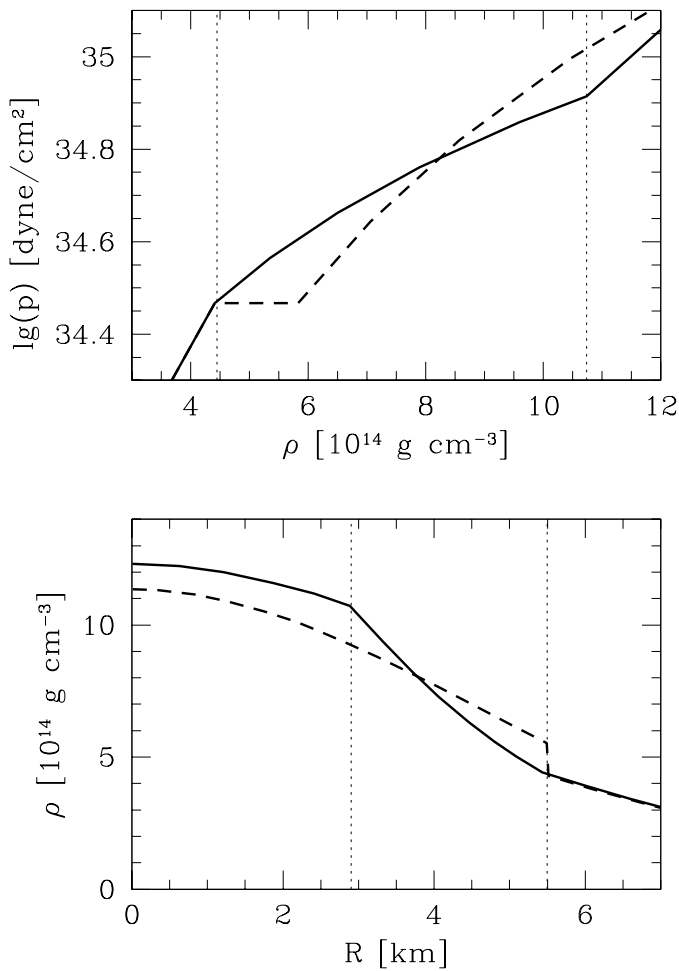}}
\caption{Two examples of EOSs employed in our calculations:
constant-pressure phase transition (dashed line, EOS with
$\Gamma_{\rm A}=\Gamma_{\rm B}=2.25$, the density jump
$\lambda=1.3$, can be found in Table~\ref{tab:firsto}) and
transition through a mixed-phase state (solid line, thin dotted
lines mark  $n_1$ and $n_2$; EOS MM of Table~\ref{tab:mixeos}). Top
panel: ${\Press}(\rho)$ diagrams. Bottom panel:
 density profiles in stellar core, for a $M=1.33~{\rm M}_\odot$
  neutron star models.}
\label{fig:EOSexamples}
\end{figure}
Studies of stability of stationary rotating configurations
require high accuracy of the 2-D calculations. This is
particularly true for the determination of a critical
configuration which separates the stable configurations from the
unstable ones. On the other hand, in our search for criteria which
link the properties of the phase transition in dense matter with
its effect on the stability of stellar configurations with
new-phase cores we wanted to explore possibly large space of the
phase transition parameters. In view of this, we decided to use
analytic forms of the EOSs. In this way, the EOSs were easy to
handle, and we could easily change the phase-transition parameters
in a continuous way, detecting with high precision the appearance
of instability. Moreover, with our numerical methods of solving
the hydrostationary equilibrium problem, we had no unnecessary
preoccupation with control of precision (see Sect.\
\ref{sec.numerical}).
\subsection{Constant pressure phase transition}
\label{sect:EOS.pure}
Normal phase is described by a polytrope of index
$\Gamma_{\rm A}$. The phase transition to a dense phase takes
place at a specific pressure, ${\rm P}_{\rm AB}$,  at which
normal phase of baryon  density $n_{\rm A}$ coexists with
dense phase of baryon density $n_{\rm B}$, with baryon density
jump $n_{\rm B}-n_{\rm A}>0$. Coexistence, equivalent to phase equilibrium,
corresponds to equality of baryon chemical potential of
both phases. The high-density
phase B is a polytrope of index $\Gamma_{\rm B}$.
 Details of the construction of the analytical
EOSs are given in the Appendix \ref{append.A}.
\subsection{Phase transition via the mixed-phase state}
\label{sect:EOS.mixed}
\begin{figure}[h]
\resizebox{\hsize}{!}{\includegraphics[]{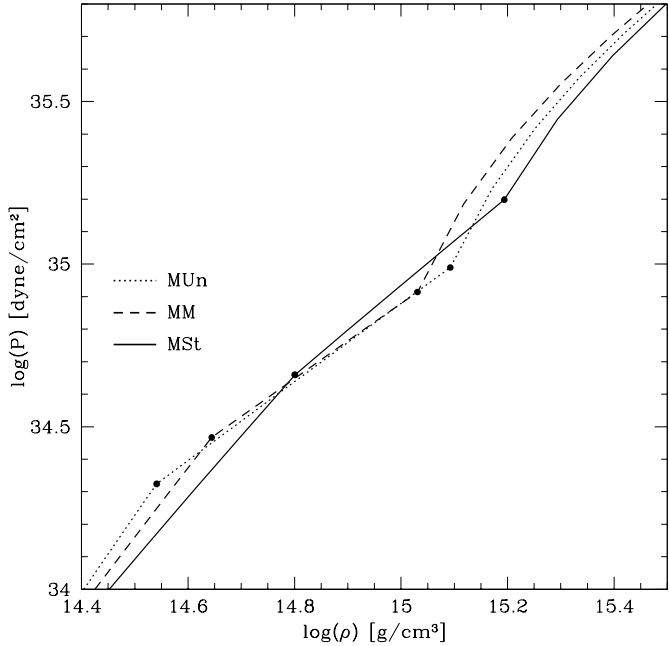}}
\caption{Examples of the three EOSs with phase transition through
the mixed-phase state, considered in  the present paper. The
parameters of the EOSs are given in Table~\ref{tab:mixeos}. }
\label{fig:mixeos}
\end{figure}

Normal phase is described by a polytrope
of index $\Gamma_{\rm A}$. Mixed phase starts at baryon density
$n_1$, and ends at $n_2$; it has a polytropic EOS with index
$\Gamma_{\rm }<\Gamma_{\rm A}$. This approximation is rather good
for a sizable lower-density part of the mixed-phase segment (see
\citealt{mix_quake2005}). To be specific, pure high-density phase,
which exists at $n>n_2$, was assumed to be quark matter, with a
bag-model EOS. Details of construction of the analytical EOSs are
given in the Appendix \ref{append.B}. Depending on the values of
parameters, we obtained three possible types of the EOS with mixed
phase region.   First, we obtained EOSs which give exclusively
stable models between the minimum and maximum allowable mass; they
are called MSt ({\bf M}ixed {\bf St}able). Second,  we obtained
the EOSs which produced two distinct stable families of neutron
stars, i.e., low density family and high density family. They are
represented by two segments in the mass-central density plane and
 are separated by a segment of unstable configurations. Such an
EOS is denoted by MUn ({\bf M}ixed {\bf Un}stable). There are also
EOSs which yield a vanishingly short unstable segment of static
configurations, so that the configurations are ``marginally stable
there'' (an inflection point). This class of EOSs will be denoted by
MM ({\bf M}ixed  {\bf M}arginally Stable). Parameters of examples of
analytical EOS of each of these three types are given in
Table~\ref{tab:mixeos}, and their $P(\rho)$ plots are shown in Fig.\
\ref{fig:mixeos}. { These examples of the EOSs have $n_1$ similar
to the models tabulated in Tables 9.2-9.3 by \citet{glendbook97}. Our
values of $n_2\sim 3n_1$ are systematically lower than those of
\citet{glendbook97} which are characterized by  $n_2\sim 4n_1$.
However, this difference is not relevant for the general results of
our analysis of stability of rotating stars. }
\begin{table}[t]
\begin{center}
\caption {Main parameters of the EOSs with mixed-phase
segment. Below the mixed-phase transition point $n_1$  a
polytropic EOS with $\Gamma_{\rm A}$ is used. Mixed phase
extends within
 $n_1 < n_{\rm b} < n_2$, and is described by a
polytrope with adiabatic index $\Gamma_{\rm m}$.
Above the density $n_2$  we assume pure quark matter with  MIT bag model
EOS $p={1\over 3}(\rho-\rho_0)c^2$. In all
cases the dimensionless polytropic pressure coefficient $K$
was equal  0.025
 (see Appendix~A for details).
$M_{\rm b,max}^{\rm stat}$ and $M^{\rm stat}_{\rm max}$ denote
the maximum allowable baryon and gravitational mass of the
 non-rotating star. The EOSs are labeled as follows: MSt
 produces a stable back bending, MUn - an  unstable one, and MM
 produces a  marginally stable case
 (for more details see the text).}
\begin{tabular}{c|c|c|c|c|c}
EOS & $\Gamma_{\rm A}$ & $n_1$
& $\Gamma_{\rm m}$ & $n_2$
& $M_{\rm b,max}^{\rm stat}~(M^{\rm stat}_{\rm max})$\\
 &    & $[{\rm fm^{-3}}]$
&   & $[{\rm fm^{-3}}]$
& $[M_\odot]$\\

\hline\hline
MSt & 2 & 0.35 & 1.5 & 0.8 & 1.508 (1.393) \\
\hline
MUn & 2.5 & 0.2 & 1.3 & 0.65 & 1.586 (1.453) \\
\hline
MM & 2.25 & 0.25 & 1.25 & 0.57 & 1.685 (1.534)
\label{tab:mixeos}
\end{tabular}
\end{center}
\end{table}

\section{Numerical methods}
\label{sec.numerical}
Hydrostationary stellar models have been computed by solving
the Einstein equations for stationary axi-symmetric space-time.
The numerical calculations were performed by means of the
{\tt rotstar} code, a part of the {\tt LORENE} public-domain
object-oriented C++ scientific library based on spectral methods.
For the complete set of equations as well as the tests and the
description of the numerical code we refer the reader to the articles
by \citet{BonazGSM93}, \citet{BonazGM98} and \citet{GourgHLPBM99}.
For the purpose of the present work,
we have employed two (in the case of a constant-pressure phase
transition) and three (phase transition via a mixed-phase state)
spectral domains to describe the stellar interior and surface.
The division into domains makes use of the adaptive coordinates
while setting the boundaries between different phases. Therefore,
the density field is smooth in each domain
 and the resulting accuracy is very high --
 in terms of GRV2 and GRV3 virial error indicators
(see \citealt{NozawSGE98}) the relative error was on average
$\sim 10^{-7}$.

\section{Back-bending and stability of rotating configurations}
\label{sec.back-bending}
As it has has been already mentioned by us in the previous paper
treating specifically the  back bending in hyperon stars
\citep{zhgb04}, the back bending phenomenon itself is strictly
connected with the existence of a minimum of $\mb$ along the
sequence of rotating configurations with fixed $f$. More
generally, the softening of the EOS leads to the flattening of the
$M(R_{\rm eq})$ and $M(\rho_c)$ curves representing these
sequences. Here, $R_{\rm eq}$ is the circumferential equatorial
radius.  The effect is larger for larger values of $f$, and it
appears above some critical value of $f$. The softening of the EOS
 results then in a local maximum of $\mb$. In such a case, there
exists a region, close to the local maximum mass, in which the
decrease of $J$ leads to the increase of $f$, and this is
equivalent to back bending. The fragment of the curve, for which
$M$ decreases as a function of $\rhoc$,  does not necessarily
correspond to the instability region -  the decrease at fixed $f$
does not imply the decrease at fixed $J$. Only the latter
condition is equivalent to the instability of rotating
configuration with respect to the axi-symmetric perturbations
 \citep{FIS88}.
Summarizing,  the particular shape of the $M(R_{\rm eq})$ curve
for a fixed  $f$ indicates the presence of the back bending
phenomenon, whereas the shape of $M(R_{\rm eq})$ at  a fixed total
angular momentum $J$ tells us if the configurations are stable.

\begin{figure}

\resizebox{\hsize}{!}{\includegraphics[]{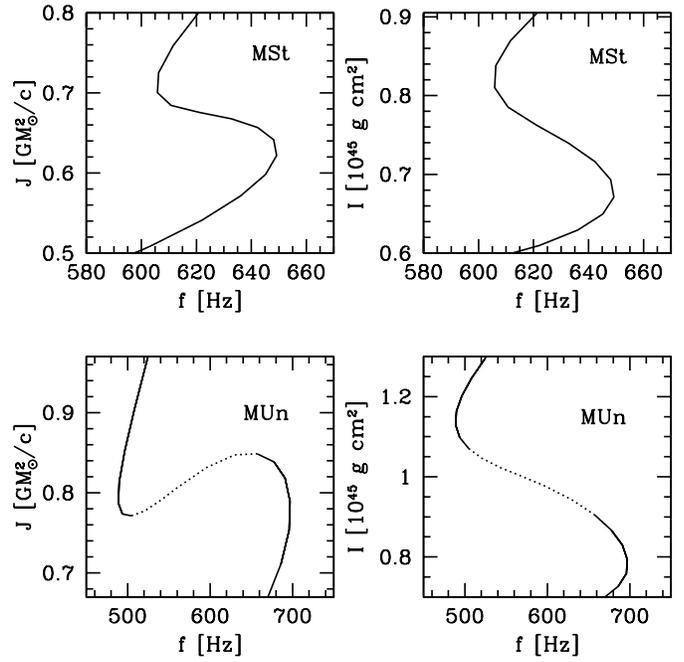}}

\caption{Total angular momentum  versus rotation frequency $f$ (left panels),
 and moment of inertia $I\equiv J/\Omega$ versus $f$ (right panels),
  for EOSs  MSt and  MUn.
The  stability criterion is easily applied to left panels. It
is clear  that for the MSt EOS back bending feature is not
associated with an instability, with all configurations being
stable. On the contrary, the  MUn EOS produces back bending
with a large segment of unstable configurations.
Simultaneously,  the $I(f)$ curves for both EOSs are very
similar and apparently show  very  similar back-bending
shapes. }
 \label{ji}
\end{figure}
%

In the present  paper we will show examples of the EOS with
constant-pressure  (density-jump) phase transitions and
 mixed-phase  transitions for which one of the two situations
 occurs:

\begin{itemize}
\item{1. all configurations are stable but the back bending
exists.}

\item{2. the phase transition results in the instability region for rotating
stars.}
\end{itemize}

To distinguish between these two cases we can look at the behavior
either of the curves $\mb(\rhoc)_{J}$ or $J(\Omega)_{\mb}$; the
quantity fixed along a sequence is indicated by the lower index.
For such curves the instability criterion directly applies.
However, is not so easy to detect the instability using the
$I(f)_{\mb}$ plot, which may be similar to that corresponding to a
fully stable sequence. An  example is presented in Fig.\ \ref{ji},
 where we plot two functions:
$J(f)_{\mb}$  and $I(f)_{\mb}$ for the two models of EOS (MSt and MUn)
which are described in detail in Sect.\ \ref{sec.mixed}. Upper
panels correspond to the stable model MSt and lower to the
MUn model for which a region of unstable configurations exists.
The difference between the MSt and MUn cases is clearly visible in
left panels ($J(f)_{\mb}$ curves), where we can easily recognize
the instability region for MUn model, by  applying the condition
${({\rm d}J/{\rm d}\rhoc)}_{\mb}>0$, and keeping in mind that
$\rhoc$ is monotonic along this curve. However, the upper and
lower right panels ($I(f)_{\mb}$) are quite similar,  without any
qualitative difference. Simultaneously,  the  back bending phenomenon in the
MUn case is very large (almost by 200~Hz),  and obviously the
phase transition results in an instability (increase of $J$ for
increasing $\rhoc$). And yet,  $I(\rhoc)$ is  a monotonic
decreasing function all the time.
\subsection{Transition to a mixed phase}
\label{sec.mixed}
\begin{figure}
\resizebox{\hsize}{!}{\includegraphics[]{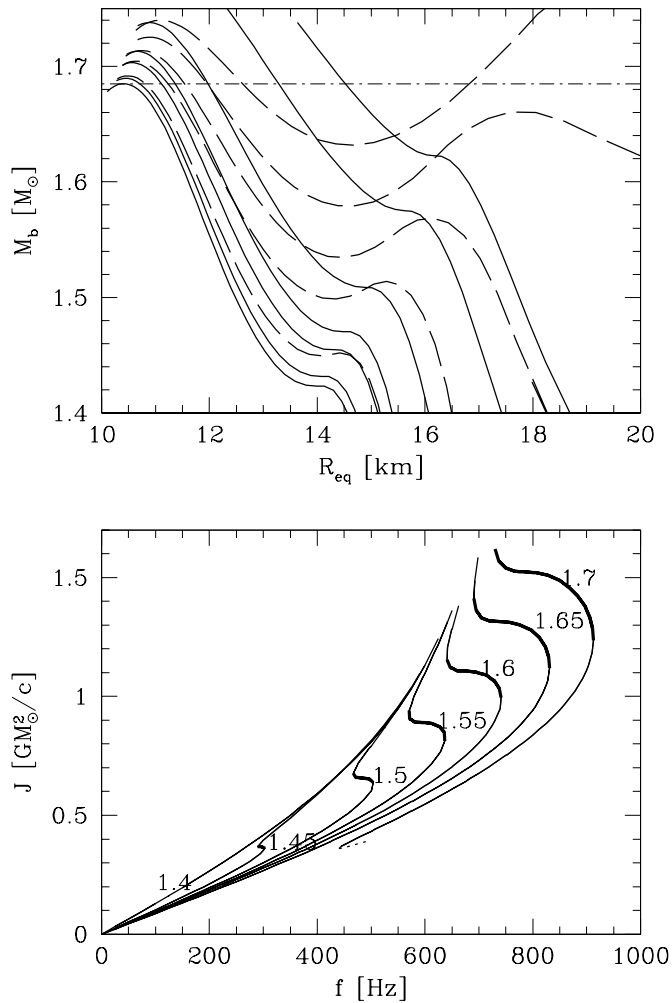}}
\caption{Upper
panel: stellar  baryon mass vs radius for fixed frequency (dashed
lines) and fixed total angular momentum (solid lines),  for the
model MM of mixed-phase EOS. Bottom panel: stellar angular momentum
as a function of the rotational frequency for fixed baryon mass
(indicated as a label, in solar masses) for the same MM EOS. This
EOS corresponds to the marginal case from the point of view of
stability -- the curves $M(R_{\rm eq})_{J}$ and $J(f)_{\mb}$ have
flat horizontal regions. The regions of back-bending are drawn by
thick lines.}
 \label{mbrmm}
\end{figure}
%
\begin{figure}
\resizebox{\hsize}{!}{\includegraphics[]{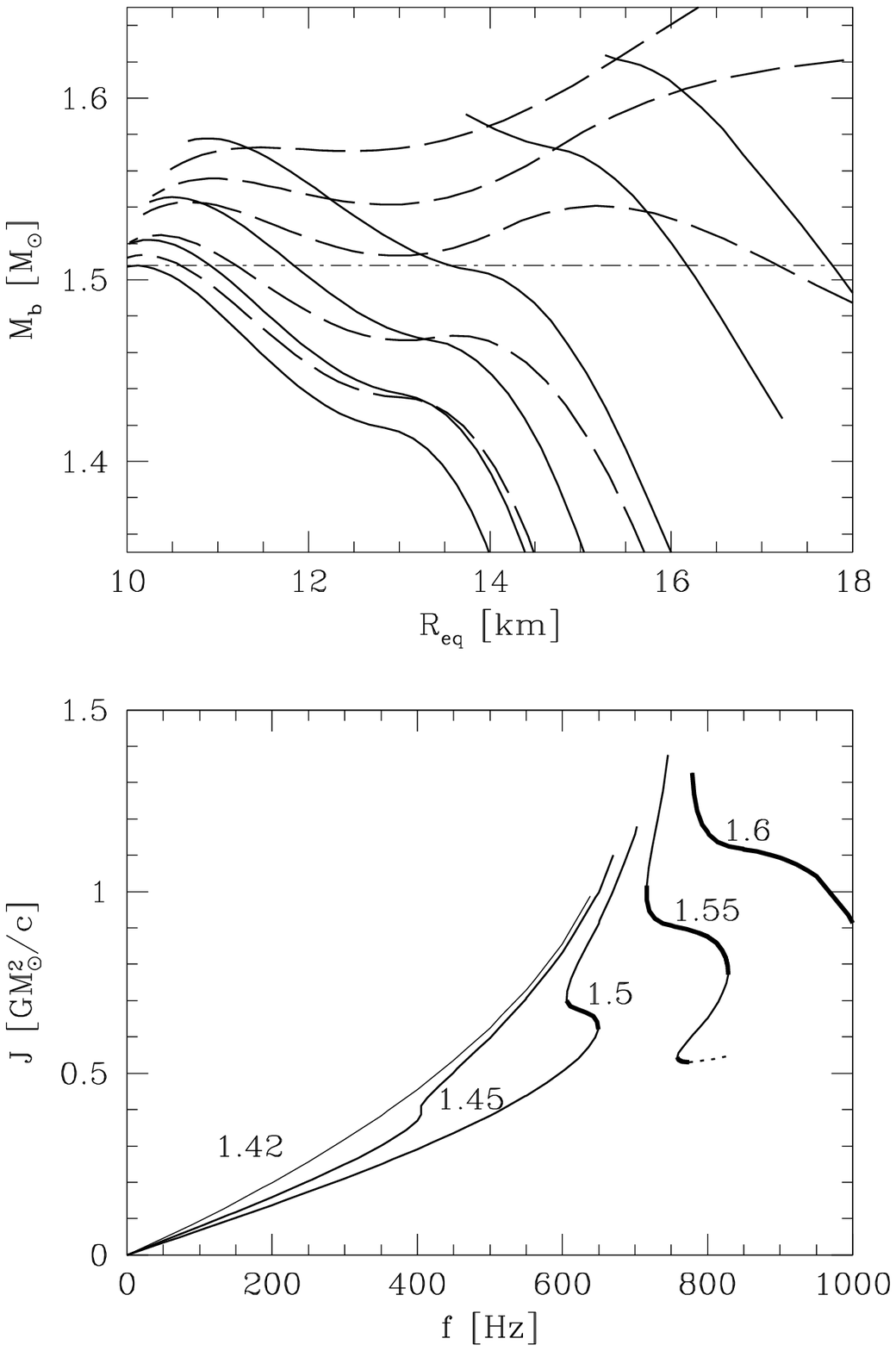}}
\caption{The same as in Fig. \ref{mbrmm} but for the  MSt EOS. Phase transition
does not result in the stability loss  -  all configurations are stable. }
 \label{mbrms}
\end{figure}
%
\begin{figure}
\resizebox{\hsize}{!}{\includegraphics[]{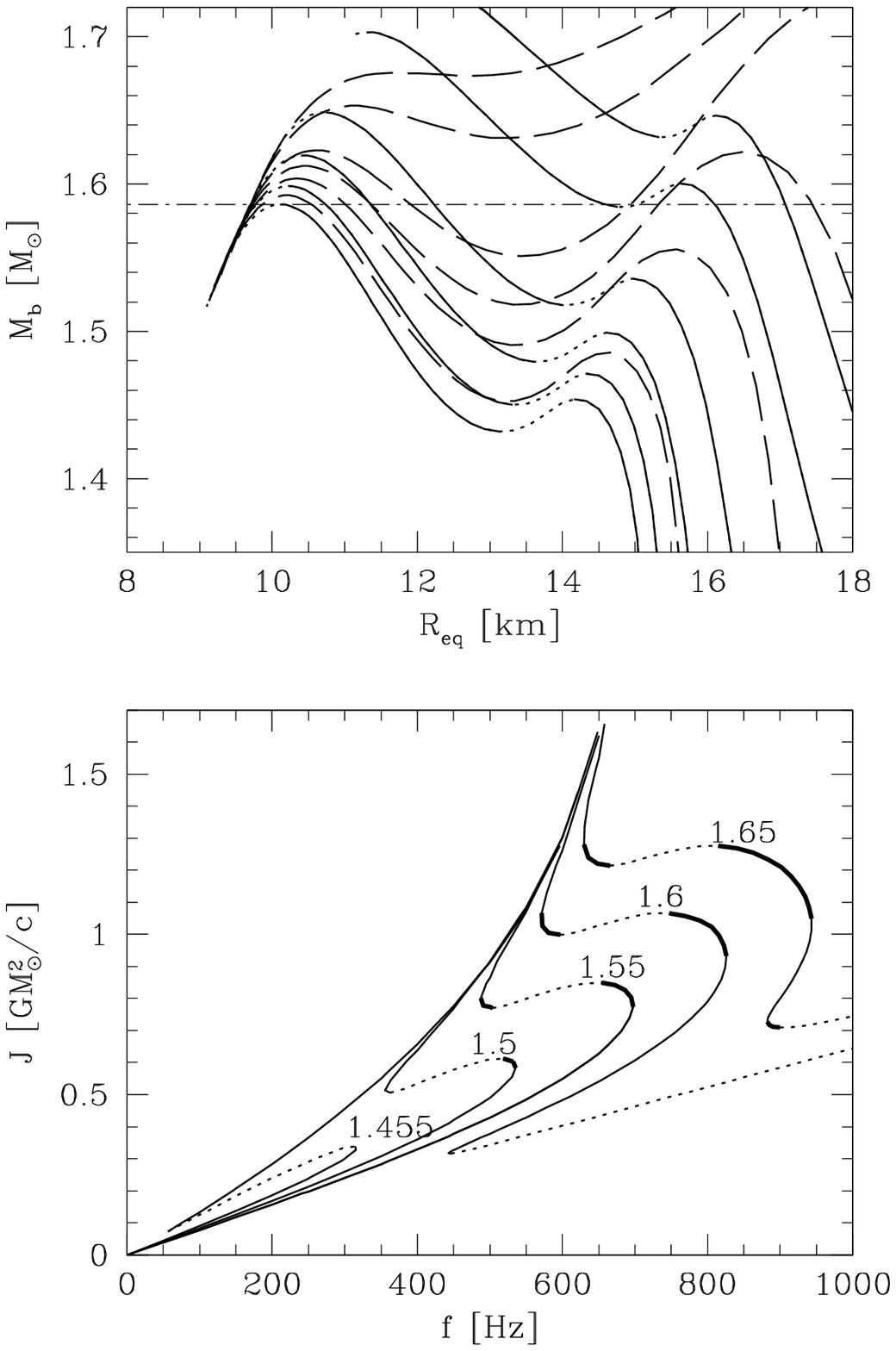}}
\caption{The same as in Fig. \ref{mbrmm} but for  MUn EOS.
Upper panel: Phase transition
 results  in the stability loss, i.e., there exists a region where $\mb$
 decreases with increasing $\rhoc$ at fixed $J$ (marked by dotted lines).
 Lower panel:
 plots in the $J-f$ plane, where on the dotted segments
  $J$ increases as  $\rhoc$ increases (unstable configurations).
 Both features indicate instability
 with respect to the axi-symmetric perturbations.}
 \label{mbrmu}
\end{figure}

In Figs. \ref{mbrmm}-\ref{mbrmu} we present
the results obtained for three different choices of the EOS's parameters
describing mixed phase phase transitions (Table~\ref{tab:mixeos}).

Model MM has been chosen to be a marginal one - there exists a
region where the curve $\mb(R_{\rm eq})$ (or $\mb(\rhoc)$) is
locally horizontal,  which means  the marginal stability. The
function $\mb(R_{\rm eq})$ for non-rotating star ($f=0$) has an
inflection point slightly above the configuration  where phase
transition to the mixed phase occurs. At a fixed $f$, the
condition for an inflection point reads,
%
\begin{equation}
\left(\frac{{\partial}\mb}{{\partial}\rhoc}\right)_{f}=0~,
~~~~~~~~\left(\frac{{\partial}^2\mb}{{\partial}\rhoc^2}\right)_{f}=0~.
\label{eq.inflat}
\end{equation}
%
 In a  special static case ($f=0$) this condition  is
 fulfilled for $\mb\simeq 1.42\ \msol$ and $R\simeq
14$~km. For  this marginal EOS model,  the back bending phenomenon
exists at
 all rotation rates, and for baryon masses larger than $1.42\ \msol$.
 The region for which back-bending
exists is bounded by the points corresponding to the local
maxima and minima of the function $\mb(R_{\rm eq})$ at fixed
rotational frequency $f$. In the marginal model for which the
inflection-point condition, Eq.\ (\ref{eq.inflat}),  is
fulfilled for  non-rotating stars, even an infinitesimally
slow rotation leads to the minimum and maximum of the function
$\mb(R)_{f}$ (albeit infinitesimally close to each other),
 which is equivalent to the back-bending (see dashed curves in
Fig.~\ref{mbrmm}). However,  imposing rotation does not result in
the instability - the functions  $\mb(\rhoc)_{J}$ are monotonously  increasing,
 independently of the rotation rate (solid curves in Fig.~\ref{mbrmm}).
Strictly speaking,  for each $J$ there exists an inflection
 point of the function  $\mb(\rhoc)_{J}$ (within the accuracy of our calculation).
The size of  back bending, defined as a frequency range for which
$f$ increases for decreasing $J$
(and increasing $\rho_{\rm c}$),  depends on the mass of the star (see bottom panel of
Fig.\ \ref{mbrmm}).
In the model MM, the back bending range varies from 0 (for $\mb \le 1.42\msol$)
to $ \sim 180$~Hz for the star with mass $\mb=1.685\msol$,
 equal to the maximum mass of non-rotating configurations
 (i.e. not supra-massive ones)

Model MSt produces a set of stable configurations in
 the region of phase transition (Fig. \ref{mbrms}). Here, $\mb$
is an increasing function of $\rhoc$ for non-rotating
configurations,  as well as along  all rotating sequences of
configurations at fixed $J$, terminating at the global maximum
mass configurations. The back-bending phenomenon is in this
case limited to the baryon masses larger than that for which
the curve $\mb(R_{\rm eq})$ (or equivalently $\mb(\rhoc)$) at
fixed $f$ has a flat, horizontal region (strictly speaking,
where
 the Eq.\ (\ref{eq.inflat}) is fulfilled).
The numerical values for the MSt model
are $\mb > 1.45\msol$ and $f>400$~Hz (see for example
the bottom panel of Fig.\ \ref{mbrms},  where the curve for $\mb = 1.45\msol$
has almost vertical part).

For the MUn model there exists a region for which configurations are
unstable, i.e. the  baryon mass is decreasing function of $\rhoc$
at fixed $J$ (Fig. \ref{mbrmu}).
In some sense this instability is not very strong - the difference
between the maximum (local) and minimum mass is of the order of
0.3\%. However this feature (existence of instability region)
is characteristic to all
rotational frequencies - for all values of angular momentum ($J$) fixed along
the curve,  the baryon mass has local maximum and local minimum  connected by
an unstable sequence of stellar configurations.
\subsection{Constant pressure phase transition}
\label{sec.maxwell}

For non-rotating configurations, the reaction of the star to a
constant pressure (first order) phase transition has been
studied in detail in the second half of 1980s (see
\citealt{zhs87} and references therein). The appearance of a
new, dense phase in the center of the star results in the
change of the derivatives of the global stellar parameters
with respect to $\rhoc$ (see the formula B6 in the appendix
of an article by \citealt{zhs87}). Two important dimensionless
parameters are: fractional density jump $\lambda=\rho_{\rm
B}/\rho_{\rm A}$ and the relativistic parameter $x_{\rm
A}\equiv P_{\rm AB}/(\rho_{\rm A}c^2)$. There exists a
critical value of $\lambda$, $\lambda_{\rm crit}={3\over
2}(1+x_{\rm A})$, such that for $\lambda>\lambda_{\rm crit}$
configurations with an infinitesimally small  B-phase core are
unstable with respect to collapse into a new configuration
with a large core of the dense phase. Putting it differently,
a phase transition with $\lambda>\lambda_{\rm crit}$
destabilizes the star at central pressure $P_{\rm c}=P_{\rm
AB}$ at which the phase transition occurs. It should be
stressed that while $\lambda<\lambda_{\rm crit}$ guarantees
stability of small-core configurations, it does not assure the
stability of configurations with a finite, or -- in an extreme
case -- a large  core. In such a case the instability would
result from the softness of the ${\rm B}$ phase somewhat above
$\rho_{\rm B}$ and not directly from an over-critical
$\lambda$. In other words the compressibility of a matter
leads to the larger mean density { in the core} than the
value $\rhob$ at the phase boundary. { The response of the
whole star to the appearance of the dense core built of the
S-phase of the matter  is determined by the mass and radius of
this core (strictly speaking, this statement is true for
non-rotating configurations, see \cite{zhs87};
for rotating ones also rotation
rate and resulting oblateness play role). As a
result } even if $\lambda<\lcrit$ the first order phase
transition can lead to the unstable configurations for finite
size of the core. As a result for the given model of the
matter in the phases ${\rm A}$ and ${\rm B}$ there exist the maximum value
of density jump $\lmax$ for which all configurations below
maximum mass are stable. Of course $\lmax \le \lcrit$ and the
difference between $\lmax$ and $\lcrit$ is larger for softer
EOS in the phase ${\rm B}$.

Numerical results for a collection of sets of EOSs with constant
pressure phase transition are collected  in
Table~\ref{tab:firsto}. The parameters presented in this table
correspond to the onset of back bending, i.e., the rotational
frequency and baryon mass for which the curve $M_{\rm b}(R_{\rm
eq})_{f}$ or $f(J)_{\mb}$ starts to have a flat region. More
precisely, at these values of frequency and mass, an inflection
point appears in the curves under consideration. We also included
parameters of those  EOSs for which all non-rotating  stars with
$M_{\rm b}<M^{\rm stat}_{\rm b,max}$ are stable.  This means that for such
EOSs the $M_{\rm b}(\rho_{\rm c})$ curve for static configurations
increases monotonically up to $M^{\rm stat}_{\rm b,max}$. The parameter
$\lambda_{\rm max}$ gives then the maximum value of the density
jump for a fixed set of other EOS parameters (adiabatic indices
$\Gamma_{\rm A}$ and
 $\Gamma_{\rm B}$, number density threshold  $n_{\rm A}$) for which
 this property of neutron stars is valid; in other words $\lambda_{\rm max}$
 corresponds to the "marginally stable" case. Increasing $\lambda$
 implies increasing softening of the EOS by the phase
 transition.  If $\lambda > \lambda_{\rm max}$,
 the phase transition leads to the existence of an
 unstable branch of the non-rotating stellar configurations.
 This unstable branch separates stable family of neutron stars with
 A-phase cores  from a second family of superdense neutron
 stars with B-phase cores: these are two distinct neutron-star
 families.
 It should be mentioned that this feature (existence of the unstable region) does not
 depend on rotation - the unstable branches exist also for rotating configurations
 (strictly speaking for any value of a total angular momentum of the star $J$ there exist
 a region with $\left(\partial\mb/\partial\rhoc\right)_{J}<0$). We have
 tested this feature (existence of or the lack of unstable regions) for very small
 departures from marginally stable case ($|{\lambda_{\rm max}-\lambda}|<0.005$).
 From numerical results it follows that if $\lambda<\lambda_{\rm max}$ all
 rotating configurations are stable (before loosing stability at maximum mass
 point) and if $\lambda>\lambda_{\rm max}$ we have two branches of stable
 configuration for rotating stars (for any $J$).

\begin{figure}
\resizebox{\hsize}{!}{\includegraphics[]{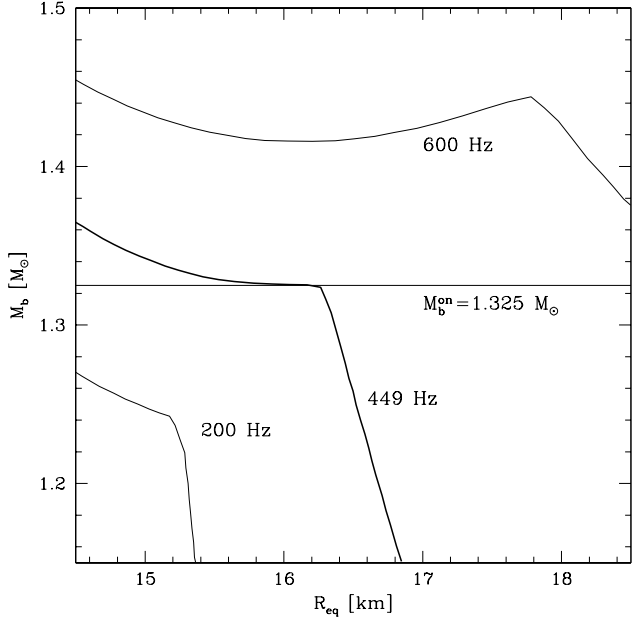}}
\caption{The definition of the frequency of the onset of
back-bending phenomenon, $f_{\rm on}$,  and corresponding
mass, $M_{\rm b}^{\rm on}$. In the presented
example
$f_{\rm on}=449$ Hz and $M_{\rm b}^{\rm on}=1.325 \msol$}
 \label{fig:bbonset}
\end{figure}

\begin{figure}
\resizebox{\hsize}{!}{\includegraphics{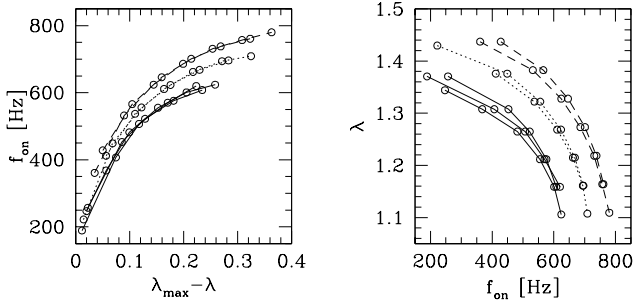}}
\caption{
The onset frequency of the back-bending phenomenon
as a function of the departure of $\lambda$ from the maximum
density jump $\lmax$ ($\lmax-\lambda$,
left panel), and density jump $\lambda$
 as a function of $f_{\rm on}$ (right panel) for models
presented in Table~\ref{tab:firsto} (solid lines for $\Gamma_{\rm A}=2$,
 dotted for $\Gamma_{\rm A}$=2.25 and dashed for $\Gamma_{\rm A}$=2.5).}
 \label{fig:onset}
\end{figure}

Picking up  the onset parameters is  visualized in
Fig. \ref{fig:bbonset} where we display $M_{\rm b}(R_{\rm eq})$
for one of the EOSs from Table~\ref{tab:firsto}. The curves are
plotted  for three frequencies, with middle one corresponding to the
back bending onset, $f=f_{\rm on}$. Last column of  Table~\ref{tab:firsto}
gives the maximum allowable baryon mass for static configurations,
$M_{\rm b,max}^{\rm stat}$. We restrict ourselves to
 back bending for the normal (non
supra-massive) stars, which appears during the spin-down evolution which terminates
eventually  by a non-rotating stable configuration. The dependence between the
back-bending onset parameters - $f=f_{\rm on}$, corresponding baryon mass, and the
intrinsic parameters of the EOS -
 the density jump $\lambda$, as well as the "departure" from the
critical configuration ($\Delta\lambda=\lmax-\lambda$)
 is presented in Fig.~\ref{fig:onset}. The three families of curves visualize
 the data from Table~\ref{tab:firsto} (solid lines for $\Gamma_{\rm A}=2$,
 dotted for $\Gamma_{\rm A}$=2.25 and dashed for $\Gamma_{\rm A}$=2.5).
 The value $\lmax$ defines the onset of back-bending at the limit $f=0$;
 in this case the back-bending phenomenon is present for any rotational
 frequency.
 As it can be seen on the left panel, the onset frequency $f_{\rm on}$
 depends very weakly on the EOS in the dense core - the main parameter describing the
reaction of the star to the appearance of this phase transition is the density jump.
The right panel presents the same data not normalized with respect to the
maximum density jump $\lmax$ - the results can be very well approximated by
the dependence $f^2_{\rm on}= a\,\Delta\lambda+b\,(\Delta\lambda)^{3/2}$.
These two plots can be treated as
 a slice through the parameter space to search for regions of the back-bending appearance -
 in the right panel, the back-bending is present, for a particular model, above
 its curve.
%


\begin{table}[h]
\begin{center}
\caption{Selected sets $\lbrace{K_{\rm A},\Gamma_{\rm
A},\Gamma_{\rm B},n_{\rm A},\lambda\rbrace}$ of EOSs with a
constant pressure phase transition. For all cases, $K_{\rm A}=0.025$
 and $n_{\rm A}=0.25$ (see the text and Appendix \ref{append.A}).}
  \begin{tabular}{l|cccc}
  EOS & $\lambda$ & $f_{\rm on}$ [Hz] & $M_{\rm b}^{\rm on}$ [$M_{\odot}$]
  & $M_{\rm b,max}^{\rm stat}$ [$M_{\odot}$] \\ \hline

\multirow{6}{0.09\textwidth}{$\Gamma_{\rm A}$=2,
                            $\Gamma_{\rm B}$=2,
                            $\lambda_{\rm max}$=1.364}
& 1.344 & 247 & 1.033 & 1.256 \\
& 1.318 & 368 & 1.071 & 1.283 \\
& 1.265 & 482 & 1.131 & 1.342 \\
& 1.212 & 555 & 1.193 & 1.407 \\
& 1.159 & 601 & 1.248 & 1.477 \\
& 1.106 & 624 & 1.286 & 1.554 \\ \hline

\multirow{5}{0.09\textwidth}{$\Gamma_{\rm A}$=2,
                            $\Gamma_{\rm B}$=2.25,
                            $\lambda_{\rm max}$=1.382}

& 1.371 & 189 & 1.021 & 1.475 \\
& 1.318 & 407 & 1.087 & 1.534 \\
& 1.265 & 507 & 1.149 & 1.597 \\
& 1.212 & 570 & 1.208 & 1.667 \\
& 1.159 & 620 & 1.272 & 1.743 \\ \hline

\multirow{5}{0.09\textwidth}{$\Gamma_{\rm A}$=2,
                            $\Gamma_{\rm B}$=2.5,
                            $\lambda_{\rm max}$=1.393}

& 1.371 & 257 & 1.034 & 1.712 \\
& 1.318 & 453 & 1.110 & 1.776 \\
& 1.265 & 522 & 1.160 & 1.844 \\
& 1.218 & 576 & 1.215 & 1.918 \\
& 1.159 & 608 & 1.258 & 1.999 \\ \hline

\multirow{6}{0.09\textwidth}{$\Gamma_{\rm A}$=2.25,
                            $\Gamma_{\rm B}$=2.25,
                            $\lambda_{\rm max}$=1.432}

& 1.376 & 412 & 1.307 & 1.642 \\
& 1.322 & 537 & 1.387 & 1.709 \\
& 1.268 & 611 & 1.463 & 1.782 \\
& 1.215 & 662 & 1.536 & 1.861 \\
& 1.161 & 694 & 1.597 & 1.947 \\
& 1.107 & 709 & 1.636 & 2.041 \\ \hline

\multirow{6}{0.09\textwidth}{$\Gamma_{\rm A}$=2.25,
                            $\Gamma_{\rm B}$=2.5,
                            $\lambda_{\rm max}$=1.444}

& 1.430 & 222 & 1.242 & 1.803 \\
& 1.376 & 449 & 1.325 & 1.859 \\
& 1.322 & 556 & 1.403 & 1.940 \\
& 1.268 & 623 & 1.477 & 2.016 \\
& 1.215 & 668 & 1.545 & 2.099 \\
& 1.161 & 696 & 1.602 & 2.190 \\ \hline

\multirow{7}{0.09\textwidth}{$\Gamma_{\rm A}$=2.5,
                            $\Gamma_{\rm B}$=2.5,
                            $\lambda_{\rm max}$=1.472}

& 1.437 & 361 & 1.559 & 1.967 \\
& 1.382 & 532 & 1.661 & 2.040 \\
& 1.329 & 624 & 1.754 & 2.119 \\
& 1.273 & 686 & 1.844 & 2.205 \\
& 1.219 & 731 & 1.933 & 2.297 \\
& 1.164 & 757 & 2.001 & 2.398 \\
& 1.109 & 780 & 2.076 & 2.509 \\ \hline

\multirow{6}{0.09\textwidth}{$\Gamma_{\rm A}$=2.5,
                            $\Gamma_{\rm B}$=3,
                            $\lambda_{\rm max}$=1.487}

& 1.437 & 428 & 1.590 & 2.348 \\
& 1.382 & 566 & 1.689 & 2.427 \\
& 1.329 & 646 & 1.871 & 2.512 \\
& 1.273 & 701 & 1.870 & 2.603 \\
& 1.219 & 738 & 1.945 & 2.703 \\
& 1.164 & 761 & 2.011 & ~2.811

\label{tab:firsto}
  \end{tabular}
  \end{center}
\end{table}

\section{Rotation and stability/instability  of normal configurations}
\label{sec.conjecture}
%
%
\begin{figure}
\resizebox{\hsize}{!} {\includegraphics[]{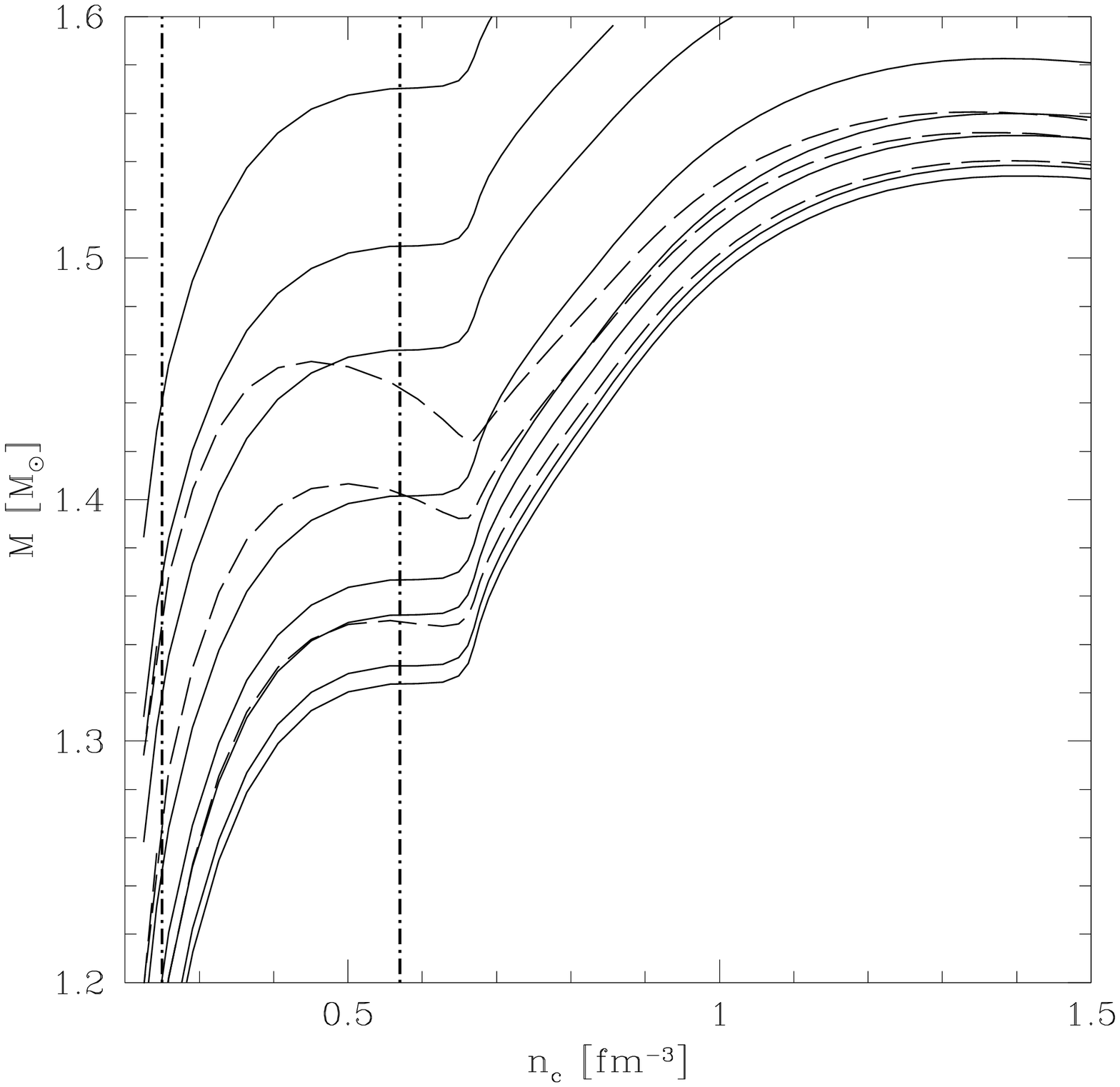}}
\caption{The mass $M$ of the star as a function of
central baryon number density $n_{\rm c}$ for the MM model.
Dashed lines - fixed rotational frequency. Solid lines - fixed
total angular momentum $J$. The vertical lines correspond to the
densities of transitions from the nuclear to mixed phase and from
the mixed phase to the pure denser phase (here - quark matter) -
see Table \ref{tab:mixeos}.} \label{mnmm}
\end{figure}
%
Rotation can influence  stability of a star of a given baryon
mass. In particular, stars with $M_{\rm b}>M^{\rm stat}_{\rm b,max}$
cannot exist without rotation and collapse into black hole as
soon as their rotation frequency falls below a certain minimum
value. Here, however, we will restrict to a different problem
of stability, which will concern the normal configurations
only.

In what follows we will use the term ``stability'' in a
restricted sense. Namely, by stability (instability) of an
equilibrium configuration we will mean  stability
(instability)  with respect to radial perturbations in the
non-rotating case, and  with respect to axi-symmetric
perturbations  for rotating configurations.

We studied  a very large set of EOSs  with  phase transitions at
constant pressure, as well as those with transition  through a
mixed phase state. We then produced static sequences and normal
rotating sequences for these EOSs. Our calculations were very
precise, because we used analytic forms of the EOSs.  The results
for both constant pressure phase transitions, and those proceeding
through mixed phase, turned out to be qualitatively  the same. In
all cases,  if non-rotating configuration were stable
(monotonically increasing $\mb(\rho_c)$ and $M(\rho_c)$), then for
any value of the total angular momentum $J$  the functions
$\mb(\rho_c)_{J}$ and $M(\rho_c)_{J}$ were monotonically
increasing, too. Thus, when all non-rotating configuration with
$M_{\rm b}<M^{\rm stat}_{\rm b,max}$ were stable (with respect to radial
perturbations), all normal rotating configurations were stable too
(with respect to axi-symmetric perturbations). On the other hand,
if for non-rotating stars there existed a region with decreasing
$\mb(\rho_c)$  and  $M(\rho_c)$, even extremely small one with a
very shallow minimum, then an unstable region persisted within the
rotating configurations, at each value of $J$. These two cases are
illustrated in  the Figs.\ \ref{mbrms} and \ref{mbrmu}.

We studied also the  case of marginally stable EOS. An
inflection point,  witnessing marginal stability,  present in
the $\mb(R)$ or  $M(\rhoc)$ curves for non-rotating stars continued to exist in
the $\mb(\req)$ or $M(\rho_c)_{J}$ curves for normal rotating stars
(Figs \ref{mbrmm}, \ref{mnmm}).

The analysis of numerical results leads us to an interesting
conclusion. Namely, for an EOS  with a phase transition
(constant pressure one or through mixed-phase state), rotation
neither stabilizes nor destabilizes normal sequences of
stationary configurations based on this EOS. We define a {\it
family} of configurations as a compact set (in mathematical
sense) of configurations. Similarly, an EOS leading to a
marginally stable point for non-rotating stars, produces also
spin-down evolution tracks with a marginally stable point. Our
result can be formulated as three  conjectures: \vskip 3mm
\parindent 0pt
 \vskip 2mm
 {\bf I}~{\it If an EOS with a phase transition  gives a single family  of
static stable neutron stars then it produces also  a single
family of stable rigidly rotating normal stars.} \vskip 2mm
\vskip 2mm
 {\bf II}~{\it If an EOS with a phase transition gives two disjoint
families of stable static stars then it gives also  two disjoint
families of stable rigidly rotating normal stars.}
\vskip 2mm
{\bf III}~{\it If an EOS with a phase transition gives two families of
stable static stars separated by a marginally stable
configuration,  then it gives also two  families of stable rigidly
rotating normal  stars separated by a line consisting of
marginally stable configurations.}
%
\section{Corequake resulting from instability}
\label{sect:corequake}
%
\begin{figure}
\resizebox{\hsize}{!}{\includegraphics[]{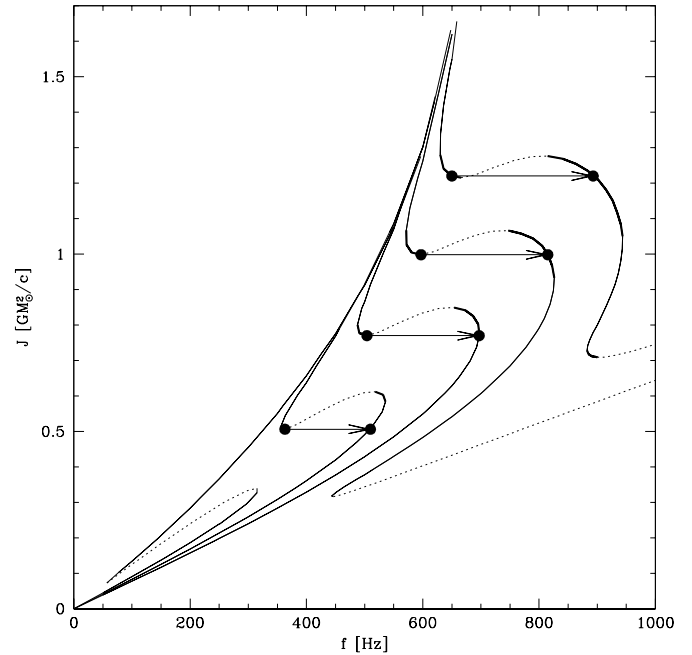}}
\caption{Evolution of an isolated pulsar loosing
angular momentum,
 after it reaches the instability region
 in $J-f$ plane and then collapses.
 Arrows lead from unstable configuration to a collapsed stable
 one, with the same baryon mass and angular momentum. Dotted lines -
 unstable configurations.
} \label{fig:arrows}
\end{figure}
%
\begin{figure}
\resizebox{\hsize}{!}{\includegraphics[]{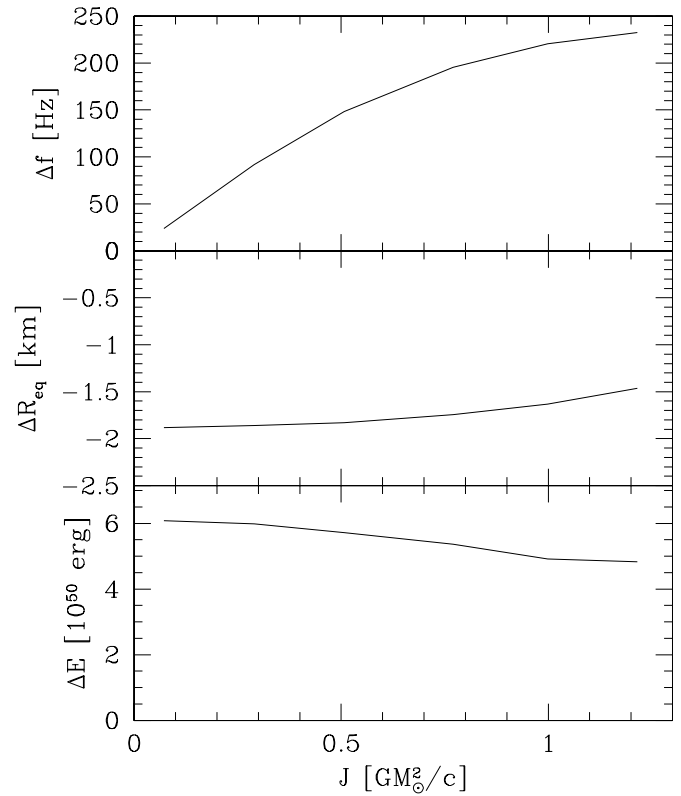}}
\caption{Changes  of stellar parameters of a rotating solitary neutron star,
due to a collapse which occurs after  a pulsar loosing angular momentum
reaches an unstable configuration.
}
\label{fig:delun}
\end{figure}
%
\begin{figure}
\resizebox{\hsize}{!}{\includegraphics[angle=-90]{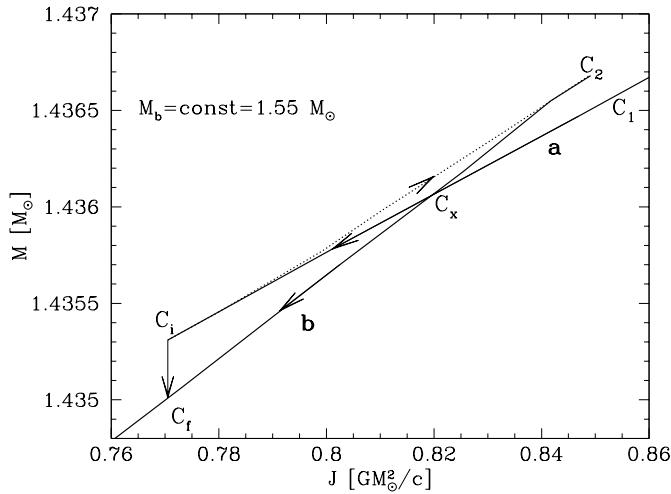}}
\caption{Total gravitational mass of the star as a function of its
angular momentum,  for fixed baryon number of the star for our EOS
model MUn. The central density is increasing along this curve as
marked by the arrows. The upper segment (dotted)
corresponds to the unstable configurations. Two cusps reflect
strict property
 that the mass and angular momentum have simultaneous
  extrema along the path with fixed baryon
number.}
 \label{jm}
\end{figure}
%
\parindent 21pt
As was shown on many occasions in the preceding sections, an
isolated neutron star, loosing its angular momentum, moves
down along  the line of fixed $\mb$ in the $J(f)$ plane, and
can at some moment reach  the {\it instability point}  (i)
( the minimum of $J(f)$
at fixed $\mb$ in  Fig.~\ref{fig:arrows}).
The subsequent behavior of the star cannot be described by our
stationary rigidly rotating model. In real world, the star has
to  collapse, rearranging the angular momentum distribution in
its interior. What we can do, is determining the final
stationary state (f), which by construction will be a stable
rigidly rotating neutron star. We assume, that the transition
conserves the baryon number of the star and is sufficiently
rapid so that the angular momentum loss can be neglected.
Therefore, the final stable configuration will have  the same
$M_{\rm b}$ and $J$ as the unstable initial one, $M_{\rm
b,f}=M_{\rm b,i}$ , $J_{\rm f}=J_{\rm i}$. The difference
between the parameters of these configurations (mass,
equatorial radius, moment of inertia) gives us the energy
release, and changes in equatorial radius and rotation
frequency,  due to the collapse implied by the
instability.

Examples  are shown  in { Figs~\ref{fig:arrows}-\ref{jm}
for our EOS model MUn}. As the star becomes more compact,
collapse is accompanied by the decrease of the equatorial
radius (by a few kilometers) and by a significant spin up. For
a given EOS, the changes in radius, energy release, and
spin-up are function of the angular momentum at the
instability point: $\Delta R_{\rm eq}(J_{\rm i})$, $\Delta
E_{\rm }(J_{\rm i})$, $\Delta f(J_{\rm i})$.  As we see in
Fig.\ \ref{fig:delun}, the energy release depends rather
weakly on the rotation of the unstable configuration (i.e., on
the value of $J_{\rm i}$). { It should be mentioned that
our MUn model is only an example of the EOS resulting in the
instability region within the hydrostatic equilibria. For EOSs
with a weaker phase transition this instability region would
be narrower and the changes of stellar parameters in the
collapse would be  smaller. However an approximate
constancy of the energy release  (i.e., its  very weak
dependence on $J=J_{\rm i}=J_{\rm f}$) seems to be a generic
property of rotating neutron stars undergoing a collapse
due to a first order phase transition.  }

In order to discuss in more detail the energy release during
collapses ${\rm i}\longrightarrow {\rm f}$, we plotted in Fig.\
\ref{jm} the gravitational mass of the star, $M$,  as a function
of angular momentum,  $J$, at  fixed baryon mass: $M=M(J)_{M_{\rm
b}}$.  Consider an initial configuration $C_1$. As the star looses
angular momentum, it moves down along line $a$, and  reaches
eventually the cusp $C_{\rm i}$ (corresponding to the value of
$J=J_{\rm i}$). To continue moving on the dotted segment $C_{\rm
i}C_2$, the star would have to gain angular momentum and energy!

As we already mentioned, the evolution of the star beyond the
instability point cannot be described by our hydrostationary model. The star
can only collapse to the final configuration $C_{\rm f}$, with the same
values of $\mb$ and $J$, i.e. along vertical arrow in Fig.\
\ref{jm}. Then, it evolves down the line $b$.

We notice a very special role played by the point $C_{\rm x}$
at which line $a$ and $b$ cross. This is a degeneracy point,
which corresponds to two very {\it different} configurations
of the same $M_{\rm b}$, $M$, and $J$. However, transitions
between these two configuration are prevented by the huge
energy barrier.

The existence of the sharp cusps  at $C_{\rm i}$ and $C_2$ on
the $M(J)_{M_{\rm b}}$ track is a very stringent test of the
precision of the numerical code: it means that mass and
angular momentum extrema (for fixed baryon mass) are reached
exactly at the same point. This property follows from the
general relativistic relation  ${\rm d}M=\Omega {\rm d}J +
\gamma {\rm d} \mb$ (\citealt{Bardeen1972}). Here,
$\Omega\equiv 2\pi f$, and $\gamma$ is the ``injection energy
per unit mass''.
 This relation  has to be strictly
fulfilled by the stationary configurations.
A graph,  analogous to Fig.\  \ref{jm},
can be plotted in the $M-\mb$ plane
for configurations with fixed angular momentum $J$. Also in
this case,  the existence of sharp cusps proves the correctness
of the numerical code.
%
\section{Phase transition and pulsar timing and age}
\label{sect:timing.age}

It has been already pointed out by  \citet{nick02}, that  the
back bending phenomenon,  resulting from the growth of a
dense-phase core, can  lead to significant difference between
the actual pulsar age, $\tau_{\rm 1}$,  and that
 inferred from the measurements of the period
$P$ and period derivative, $\dot{P}$, and denoted $\tau_2$.
The calculation of $\tau_2$ is based on quite strong assumptions. Firstly,
pulsar kinetic energy loss due to radiation is given by the magnetic
dipole formula. Secondly,  non-relativistic approximation is used,
with pulsar kinetic energy given by ${1\over 2}I\Omega^2$, where
$\Omega\equiv 2\pi/P$ and pulsar moment of inertia is constant,
 independent of $\Omega$.

Following \citet{nick02}, we will use general relativistic
notion of total pulsar energy, $Mc^2$. Then the pulsar energy
balance is
%
\begin{equation}
{{\rm d}M\over {\rm d}t} = -{\kappa\over c^2} \Omega^{\alpha}
\label{eq:Erot.loss.GR}
\end{equation}
%
where the right-hand-side is the pulsar energy loss rate via
radiation of electromagnetic waves and particles.

Let us consider the increase of stellar energy due to a spin up
to frequency $f$ at constant $M_{\rm b}$. In general relativity, the
increase is given by $\Delta M(f)=[M(f)-M(0)]_{M_{\rm b}}$.
In the standard model, we neglect the effect of $f$ on stellar
structure, so that $\Delta M={1\over 2}I (2\pi f)^2$.
This is a good approximation when rotation is
slow and EOS is smooth (no phase transition).
Equation (\ref{eq:Erot.loss.GR}) can be then rewritten as
%
\begin{equation}
\dot{\Omega} = -{\kappa\over I_0} \Omega^{\alpha -1},
\label{eq:Iconst}
\end{equation}
where $I_0\equiv I(0)=const$.

\begin{figure}
\resizebox{\hsize}{!}{\includegraphics[angle=-90]{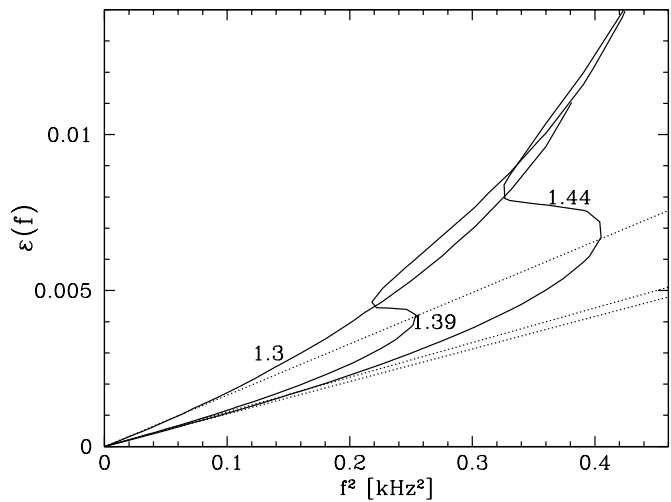}}
\caption{The relative mass-energy increase due to rotation of
the star at fixed baryon mass, $\varepsilon(f)\equiv
[(M(f)-M(0))/M(0)]_{M_{\rm b}}$,  for the  EOS with a phase
transition (MM model), is shown using solid line. Three solid
lines are labeled by the gravitational mass of the
non-rotating configuration (in solar masses). Dotted lines
correspond to the $\varepsilon(f)$ curves calculated for  the
standard model, Eq.\ (\ref{eq:Iconst}), with $\alpha=4$.
  }
\label{fig:dmf}
\end{figure}

\begin{figure}
\resizebox{\hsize}{!}{\includegraphics[]{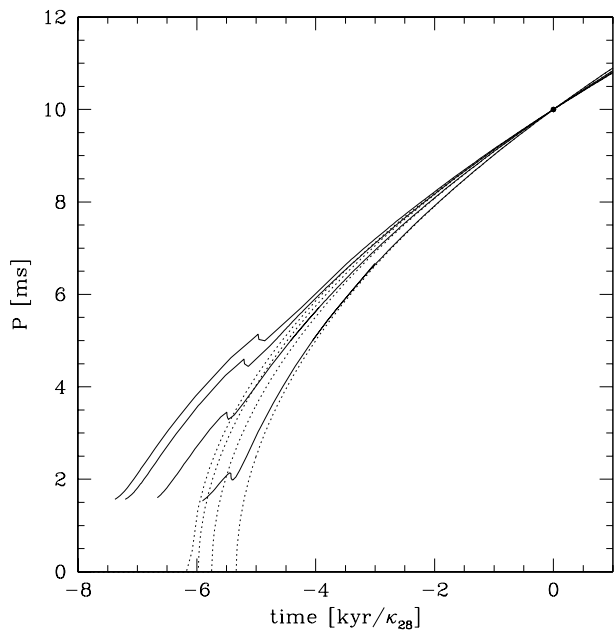}}
\caption{The evolution of the pulsar period when the energy
loss is described by the magnetic dipole braking with
$\alpha=4$. Solid curves - results for our model MM with
different values of baryon mass.
Dotted lines correspond to the standard model, Eq.\ (\ref{eq:Iconst}).
The unit of time (horizontal axis) is $1000\,{\rm yrs}/\kappa_{28}$ where $\kappa_{28}=
\kappa/10^{28}$ [cgs]
is parameter entering Eq. (\ref{eq:Erot.loss.GR})
  }
\label{fig:tt}
\end{figure}

In the case when angular momentum loss leads to
a phase transition at the stellar
center, the situation is much more complicated, because
of the strong $f$-dependence of the pulsar structure in
the vicinity of the phase transition. This difference
 is illustrated in Fig.\ \ref{fig:dmf}, where we plotted
 the quantity  $\varepsilon(f)\equiv
 \Delta M(f)/M(0)$ resulting from
 our calculations, and compared it with  results
 given by standard non-relativistic model with constant $I$.
%
As it can be seen in Fig.\ \ref{fig:dmf},  the approximation given
by Eq.\ (\ref{eq:Iconst}) significantly {\it overestimates} the
change of frequency associated with a given energy loss. This in
turn can lead to an  {\it underestimation} of the age of the
pulsar --  an  example is presented in Fig.\ \ref{fig:tt}. There,
we plot  the pulsar period, $P$,  as a function of time, obtained
by the integration of Eq.\ (\ref{eq:Erot.loss.GR}) for magnetic
dipole braking ($\alpha=4$). We also show results inferred from
the observation of a  10 ms pulsar, assuming four selected values
of the pulsar baryonic mass. Taking $P=10$ ms and corresponding
$\dot P$ the extrapolation backward in time using standard model,
Eq.\ (\ref{eq:Iconst}), diverges from exact results as soon as
$P<5$ ms. If the pulsar is born with period of  2 ms, then its
real age is significantly longer than  $P/2\dot{P}$.
%
%
\section{Discussion and conclusions}
\label{sect:conclusions}
%
The effect of phase transitions in dense neutron-star cores on
the spin-down evolution of isolated stars was studied using
analytical representations of EOSs. This guaranteed a very
high precision of our 2-D calculations and allowed us to study
a very large parameter space as far as the phase transitions
themselves are concerned. In this way, we studied a very broad
class of constant-pressure phase transitions associated with a
density jump in the EOS. We also studied the case of
transitions through a mixed-phase state.

We limited ourselves to the case of normal rotating
configurations, which are connected with the non-rotating ones by
the angular momentum loss.  We considered two types of
instabilities which bound the sets of stationary configurations:
mass shedding and instability with respect to the axisymmetric
perturbations. The EOSs split into two sets: those producing a
single family of stable stationary configurations (static {\it
and} rotating) of neutron stars and those producing two disjoint
families of stable stationary configurations. Conjectures
concerning normal configurations based on EOSs with a phase
transition has been formulated. If an EOS yields two disjoint
families of static configurations containing ``twin neutron
stars'' of the same baryon mass but different radius, then also
stable normal rotating configurations form two disjoint regions in
the mass -- equatorial radius plane, which contain ``twin neutron
stars'' of the same baryon mass but of different compactness.

Very often, neutron stars are called a second family of compact
stars, the first family being composed of white dwarfs. Therefore,
our conjecture means that an EOS which produces a third (disjoint)
family of static compact stars, produces also a third (disjoint)
family of rotating normal stars.

We have also shown the existence of a very special class of
``fine tuned EOSs'' with phase transitions which produce
marginally stable stationary configurations of normal neutron
stars, which form a boundary separating stable stationary
configurations (a line in the mass-equatorial radius plane).

Conditions on the density jump in constant-pressure phase
transitions were derived, under which their  presence in the
EOS  produces the back bending phenomenon in the spin down
evolution.

The case when a spinning down normal neutron star reaches an
unstable configuration was studied in detail. The instability
leads to neutron star collapse, associated with an energy
release in a ``corequake'', decrease of radius, increase of
central density, and spin up of the star. We have shown that
the energy release associated with such a ``corequake''
depends rather weakly on the initial  rotation frequency at
the instability point. In   our examples,  energy release was
of the order of a few times $10^{50}~$erg.

In the present paper we put accent on the numerical precision
and mathematical strictness. We  hope that in this way we
prepared  ground for further  studies of the impact of the
phase transitions in dense matter on the structure, evolution,
and dynamics of rotating neutron stars. These further studies
will be performed using realistic EOSs available in the
literature and taking into account important microscopic
aspects of the phase transitions.  The kinetics of the phase
transition coupled with stellar spin-down, and the ensuing
neutron star corequake are now being studied. These topics
will be the subject of our subsequent papers.
%
\acknowledgements{This work was partially supported by the
Polish MNiI grant no. 1P03D-008-27 and by the PAN/CNRS LEA
Astro-PF.}
%
\appendix
\section{Analytical equations of state}
\label{append:relpoly}
The relativistic polytrope, following \citet{tooper65},
relates pressure $p$ to baryon number density $\nb$ by
\begin{equation}
\label{poly_Pnb}
p({\nb}) = K{\nb}^{\Gamma},
\end{equation}
where $\Gamma$ is the adiabatic index.
The second coefficient, $K$, is often called the pressure coefficient
\footnote{Unless otherwise mentioned, the coefficient $K$ will be
measured in  $\hat{\rho}c^2/\hat{n}^{\Gamma}$ units,
where $\hat{\rho}:=1.66\times10^{14}~{\rm g/cm^3}$,
and $\hat{n}:=0.1~{\rm fm^{-3}}$.}.

Dense matter is strongly
degenerate, so that the $T=0$ approximation is valid.
First Law of Thermodynamics implies then expression for energy
per baryon
\begin{equation}
{\Ener}/{\nb} =
\frac{K{\nb}^{\Gamma-1}}{\gamma-1}+{\cal C}~,
\label{poly_ener1}
\end{equation}
where the constant ${\cal C}$ is set, for ${\nb}=0$, to be the
unit baryon rest energy\footnote{Following many authors,
we put unit baryon mass equal to mass per nucleon in the
ground state of atomic matter at zero pressure,
which is $^{56}{\rm Fe}$ crystal, $m_0=1.66\times10^{-24}~{\rm g}$}
 ${\cal C}={\m0}c^2$.

The energy density $\Ener$ is thus given by
\begin{equation}
\label{poly_ener2}
{\Ener}({\nb}) = {K\over\Gamma-1}
\nb^{\gamma} + {\m0}c^2{\nb}~.
\end{equation}
The baryon chemical potential ${\mu}$ is therefore
\begin{equation}
\mu({\nb}) = {p+{\Ener}\over{\nb}} =
{\Gamma K\over\Gamma-1}{\nb}^{\Gamma-1} + {\m0}c^2~.
\label{chem_pot}
\end{equation}
\subsection{Constant pressure phase transition}
\label{append.A}
\begin{figure}[h]
\centering \resizebox{3.5in}{!}
{\includegraphics[clip]{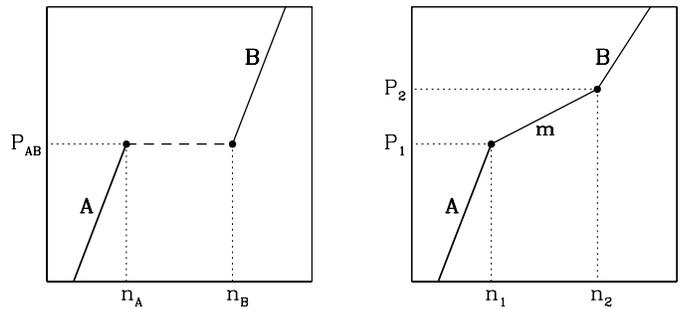}}
\caption{Examples of phase transitions considered in the text;
constant pressure phase transition (left), and the phase transition
through the  mixed-phase state (right).}
\label{fig:eos_examples}
\end{figure}

We assume that the phase transition stakes place in thermodynamic
equilibrium. In a simplest case
the transition from less dense pure phase
${\rm A}$ to high-density pure phase ${\rm B}$
occurs at constant pressure. Example of such
transition is shown on Fig.~\ref{fig:eos_examples}.
Pure phases, ${\rm A}$ and ${\rm B}$, will be approximated by polytropes,
with adiabatic indices $\gamA$
and $\gamB$, respectively. Let us also fix the mean baryon mass $\mA$
and coefficient $\KA$, and demand that the phase transition
occurs between the baryon densities $\nA$ and $\nB$, at  constant
pressure ${\rm P}_{\rm AB}$ and chemical potential $\mu$. From constancy
of pressure and chemical potential we get
\begin{eqnarray}
\label{const_press_phtr}
{\KB} & = & {\KA}\cdot {\nA}^{\gamA}/{\nB}^{\gamB}~, \\
{\mB} & = & {\mA}+ {\rm P}_{\rm AB}~\left(\frac{\gamA}{{\nA}(\gamA-1)}
-\frac{\gamB}{{\nB}(\gamB-1)}\right)~.
\end{eqnarray}

\subsection{Transition through the mixed-phase state}
\label{append.B}
Lower-density phase ${\rm A}$ extends up to pressure ${\rm P}_1$ and
density $\nb=n_1$, then follows mixed m-phase with volume fraction of
dense phase $\chi_{\rm B}$ increasing monotonically from 0 at
$n_1$ to 1 at $\nb=n_2$ and $p={\rm P}_2$. Again, for
simplicity, it will be assumed that the phase ${\rm A}$ and
mixed-phase $m$ can be approximated by polytropes (not a bad
approximation, see \citealt{mix_quake2005}). The parameters
$K_{\rm m}$, $\Gamma_{\rm m}$, and ${\mm}$ (the {\it mean}
particle mass in the mixed-phase) must be related to those of
the ${\rm A}$-phase in such a way that pressure and baryon chemical
potential stay continuous across the ${\rm A}\longrightarrow
{\rm m}$
 transition point at $n_1$,
\begin{eqnarray}
\label{mix_phase_phtr}
{\Km} & = & {\KA}\cdot{n_1}^{{\gamA}-{\gamm}}~, \\
{\mm} & = & {\mA}-{{{\rm P}_1}\over{{n_1}c^2}}
\cdot{{{\gamA}-{\gamm}}\over{({\gamA}-1)({\gamm}-1)}}~.
\end{eqnarray}
We will  further assume that the high-density phase ${\rm B}$,
existing at $n_{\rm b}>n_2$, $p>{\rm P}_2$, and $\rho>\rho_2$,
 is pure quark matter, with MIT bag model EOS \citep{jlz00},
\begin{equation}
\label{quark_simple_eos}
{p}({\Ener}) = {{1}\over{3}}({\Ener}-{\Ener}_0),~~~
{\nb}(p) = n_{0}(1+{p}/{\Ener})^{3/4}~,
\end{equation}
where ${\Ener}_0=\rho_0 c^2$ and $n_0$ are the mass-energy density
and the baryon density of the quark matter at zero pressure.
The baryon chemical potential of the B-phase is then equal to
\begin{equation}
\label{quark_chem_pot}
\mu({\Press}) = \mu_0(1+4{\Press}/{\Ener})^{1/4}~,
\end{equation}
where $\mu_0={\Ener}_0/n_0$.

From the continuity  of pressure, baryon density, and energy density
at the ${\rm m}\longrightarrow {\rm B}$ transition point, we get
\begin{eqnarray}
\label{quark_parameters} {\Ener}_0 & = & {\Ener}_2-3{\rm
P}_2({n_2})~, \nonumber \\
n_0 & = & n_2/(1+4{\rm P}_2/{\Ener}_2)^{3/4}~.
\end{eqnarray}


\end{document}